\documentclass[aps,prl,reprint,twocolumn,showpacs,floatfix,superscriptaddress]{revtex4-2}

\usepackage{amssymb,amsmath,amstext}                
\usepackage{graphicx}                                               
\usepackage{epstopdf}                                               
\usepackage{color}                                                     
\usepackage{bm}                                                        
\usepackage{appendix}                                              
\usepackage[utf8]{inputenc}
\usepackage{dsfont}                                 
\usepackage[normalem]{ulem}

\usepackage[shortlabels]{enumitem}

\usepackage{verbatim}
\usepackage{latexsym}
\usepackage{xcolor}
\usepackage{braket}
\usepackage{ulem}
\definecolor{lblue}{RGB}{51,71,158}
\usepackage[colorlinks=true,citecolor=blue,linkcolor=blue,urlcolor=lblue]{hyperref}

\DeclareMathOperator{\Tr}{Tr}

\makeatletter\AtBeginDocument{\let\@elt\relax}\makeatother

\begin{document}

\title{Many-Body Mobility Edge in Quantum Sun models}

\author{Konrad Pawlik}
\email{konrad.pawlik@doctoral.uj.edu.pl}
\affiliation{Szkoła Doktorska Nauk Ścisłych i Przyrodniczych, Uniwersytet Jagielloński, ulica Stanisława \L{}ojasiewicza 11, PL-30-348 Krak\'ow, Poland}
\affiliation{Instytut Fizyki Teoretycznej,  Wydział Fizyki, Astronomii i Informatyki Stosowanej, Uniwersytet Jagielloński, ulica Stanisława \L{}ojasiewicza 11, PL-30-348 Krak\'ow, Poland }

\author{Piotr Sierant}
\affiliation{ICFO-Institut de Ci\`encies Fot\`oniques, The Barcelona Institute of Science and Technology, Av. Carl Friedrich
Gauss 3, 08860 Castelldefels (Barcelona), Spain}

\author{Lev Vidmar}
\affiliation{Department of Theoretical Physics, J. Stefan Institute,  SI-1000 Ljubljana, Slovenia}
\affiliation{Department of Physics, Faculty of Mathematics and Physics, University of Ljubljana, SI-1000 Ljubljana, Slovenia\looseness=-1}

\author{Jakub Zakrzewski}
\email{jakub.zakrzewski@uj.edu.pl}
\affiliation{Instytut Fizyki Teoretycznej,  Wydział Fizyki, Astronomii i Informatyki Stosowanej, Uniwersytet Jagielloński, ulica Stanisława \L{}ojasiewicza 11, PL-30-348 Krak\'ow, Poland }
\affiliation{Mark Kac Complex
Systems Research Center, Jagiellonian University in Krakow, PL-30-348 Krak\'ow,
Poland. }

\date{\today}

\begin{abstract}
 The quantum sun model is an interacting model that exhibits sharp signatures of ergodicity breaking phase transition. Here, we show that the model exhibits a many-body mobility edge. We provide analytical arguments for its existence, complemented by the state-of-the-art numerical simulations analysing gap ratios, Thouless times as well as entanglement entropy of eigenstates. We also introduce the  quantum sun model with particle number conservation, and we argue that it shares many similarities with his unrestricted predecessor. 
\end{abstract}

\maketitle

\paragraph{Introduction.} 
In recent years, there has been a rapid change of the perspective on  
Many body localization (MBL) \cite{Gornyi05, Basko06, Oganesyan07, Pal10} proposed as a robust mechanism leading to breakdown of quantum thermalization in isolated many-body systems \cite{Dalessio16} in the presence of disorder and interactions. Intensive theoretical studies (for reviews see e.g.~\cite{Nandkishore15, Alet18, Abanin19}) supplemented by experiments \cite{Schreiber15,Smith16,Luschen17,Lukin19,Kohlert19, Leonard23}  
yielded a picture of MBL as of a dynamical phase of matter in which the local integrals of motion \cite{Serbyn13b, Huse14,Ros15, Thomson18} preserve information about the initial state, the transport is suppressed \cite{Nandkishore15, Znidaric16}, and the entanglement spreads slowly \cite{DeChiara06, Znidaric08, Bardarson12,Serbyn13a,iemini2016signatures}.
Disordered Heisenberg spin-1/2 chain served as one of the paradigmatic models in those investigations of MBL
\cite{Santos04, Oganesyan07, Berkelbach10, Serbyn15, Agarwal15, Bera15,Luitz16, Luitz16b, Bera17, Colmenarez19, Serbyn16,Bertrand16,Khemani17a, Serbyn17, Kjall18, Doggen18, Doggen19, Weiner19, Sierant19b, Sierant20, Mace19b, TorresHerrera20b, TorresHerrera20, Schiulaz19, Hopjan21}. 
Hence, it came as a surprise when \cite{Suntajs20e} pointed out that numerical results for this model could be interpreted in terms of a slow-down of quantum thermalization with increasing disorder strength rather than in terms of a transition to an MBL phase, challenging the very existence of MBL as of a dynamical phase of matter.
Subsequent works \cite{Sierant20b, Abanin21} showed that the existence of MBL is not directly ruled out by the arguments of \cite{Suntajs20e},  arguing that a convincing numerical demonstration of MBL~\cite{Panda20} is beyond reach of present day supercomputers~\cite{Pietracaprina18,Sierant20p}.
Those results sparkled a debate about the status of MBL~\cite{Sierant20p, Suntajs20, Laflorencie20, Sels21a, Morningstar22, Sels22}, during which 
the role of slow, but apparently persistent approach towards thermal equilibrium was emphasized~\cite{Kiefer20,Kiefer21, Luitz20, Ghosh22, Sierant22, Sierant23, Szoldra23}.


\begin{figure}[b!]
\centering
\includegraphics [width=0.45 \textwidth,angle=0]{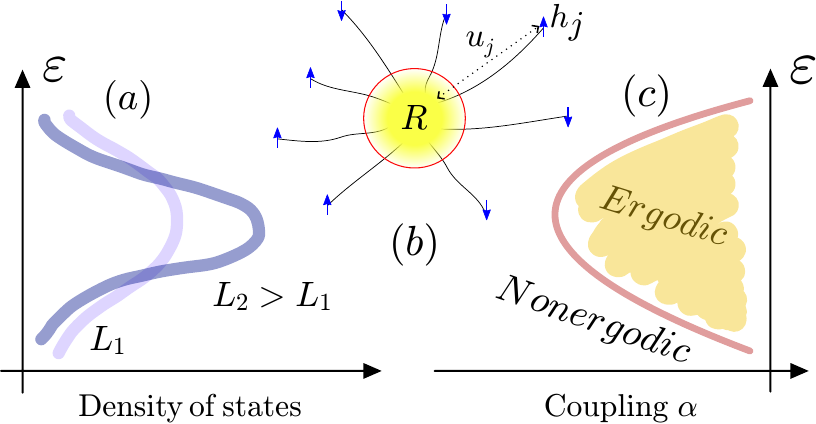}
\caption{
(a) Many-body density of states (DoS) as a function of energy density $\varepsilon$.
By increasing the system size $L$, the DoS gets exponentially suppressed away from the middle of the spectrum. (b) Visualization of the model described by  \eqref{eq:sun} and \eqref{eq:cons_model}. Particles inside the sun are described by GOE matrix $R$, while the outside spin $j$  is perturbed by the magnetic field $h_j$ and coupled to one of the spins of the sun with distance, $u_j$, dependent coupling. (c) The many-body mobility edge is manifested as a variation of the critical coupling $\alpha$ with $\varepsilon$.
}\label{fig:sketch}
\end{figure}

Recently, it was shown~\cite{Suntajs22} that 
 a  clear numerical demonstration of ergodicity breaking phase transition (EBPT) is possible for the "quantum sun" model (QSM), a model that was originally investigated in a context of quantum avalanches \cite{DeRoeck17, Luitz17} seeded by rare ergodic regions in spin chains~\cite{Dumitrescu17, Goremykina19, Dumitrescu19,Herviou19, Szoldra21}.
The QSM consists of a small delocalized core, described by a random matrix from Gaussian orthogonal ensemble (GOE) \cite{Haakebook} coupled to a chain of spins by interaction which decays exponentially with the distance, c.f. Fig.~\ref{fig:sketch}(b). The position of EBPT can be approximated~\cite{Luitz17, Suntajs22, hopjan_23} due to a simple analytic hybridization condition~\cite{DeRoeck17} which studies the scaling properties of the interaction matrix element relative to the level spacing that is the inverse of the density of states (DoS).

In this Letter, we argue that the hybridization condition does not only enable a sharp prediction of the EBPT point in QSM, but it also implies an existence of a mobility edge separating the spectrum of  ergodic and localized eigenstates. Mobility edges exist for noninteracting systems~\cite{Delande14, Pasek17, Luschen18, Li20} and they were postulated also for MBL \cite{Altshuler97Lifetime,Kjall14,Luitz15,Mondragon15, Villalonga18, Enss17,Chanda20m}, but at the same time there are theoretical arguments \cite{DeRoeck16} against their existence in interacting systems. Therefore, deciding whether many-body mobility edges are stable, or whether they are only a transient feature that disappears in the thermodynamic limit, is one of the key problems in non-equilibrium physics of disordered many-body systems.
For the QSM studied here, the
 key insight is that the DoS is  exponentially suppressed away from the middle of the spectrum, as sketched in Fig.~\ref{fig:sketch}.
We demonstrate emergence of the many-body mobility edge in the original QSM, which does not conserve the total magnetization $\hat S_{\rm tot}^z$.
We then introduce another variant of the QSM that conserves $\hat S_{\rm tot}^z$ and we argue that both models exhibits very similar properties, verifying  that the stability of EBPT in QSM is not affected by the presence of additional constants of motion.

\paragraph{Analytical arguments for the mobility edge.}
The condition for the EBPT in the QSM of size $L$ is based on the hybridization condition~\cite{DeRoeck17, Suntajs22}
\begin{equation} \label{eq:hybr}
{\cal G} \propto \alpha^L \sqrt{\rho} = {\rm const.} \;\;\; \rightarrow \;\;\; \alpha_c{\propto} \rho^{- \frac{1}{2L}}\;,
\end{equation}
where  $\alpha$ is the amplitude of interaction matrix element and $\rho$ is the DoS.
Usually, one considers the middle of the spectrum ($E \approx 0$) for which the DoS for spin-1/2 particles scales as $\rho(E=0) \propto 2^L$, giving rise to the transition point $\alpha_c = 1/\sqrt{2}$. In contrast to the extended disordered spin chains, the presence of EBPT at $\alpha \approx \alpha_c = 1/\sqrt{2}$ in QSM can be demonstrated numerically in a convincing fashion, as shown in \cite{Suntajs22} with a few indicators of ergodicity breaking.

In many-body systems with local interactions, the DoS 
{is well approximated by a Gaussian shape with possible model dependent deviations \cite{French70,Bohigas71,Hartmann04,Hartmann05}}
\begin{equation}
    \rho(E) = 2^{L} \frac{1}{\sqrt{2 \pi \sigma_E^2}} e^{-\frac{E^2}{2\sigma_{E}^2}},
    \label{eq:spacing}
\end{equation} 
where $\sigma_E$ is the standard deviation of energy, which has a leading scaling $\sigma_E =  b \sqrt{L}$. Expressing the energy $E$ in terms of the rescaled energy $\varepsilon=(E-E_{\mathrm{min}})/(E_{\mathrm{max}}-E_{\mathrm{min}})$, where $E_{\mathrm{min}}$ ($E_{\mathrm{max}}$) is the ground state (highest excited state) energy, and using the system size dependence of $\sigma_E$ and extensivity of energy, $E_{\mathrm{max} }-E_{\mathrm{min}}=aL$, we find the leading exponential scaling in \eqref{eq:spacing} is $(2 e^{ -a^2 (\varepsilon-1/2)^2/(2b^2)} )^{L}$  with constants $a$ and $b$ that are dependent on the details of the model. 
Using this scaling in the criterion \eqref{eq:hybr}, we find the position of the transition between ergodic and localized phases at rescaled energy $\varepsilon$ as
\begin{equation}
\alpha_c(\varepsilon)=\frac{1}{\sqrt{2}}e^{\frac{a^2( \varepsilon-1/2)^2}{4 b^2}}\;.\label{eq:analytic_edge}
\end{equation}
This formula reproduces the analytic result $\alpha_c(\varepsilon)=\frac{1}{\sqrt{2}}$ at $\varepsilon = 1/2$ and predicts the shape of the many-body mobility edge for any $\varepsilon \in [0,1]$. The critical assumption of the formula~\eqref{eq:analytic_edge} is that the interaction matrix element scales with distance $L$ as 
$\alpha^L$. The criterion \eqref{eq:analytic_edge} arises due to a competition of DoS with the interaction matrix element. In the following, we start by investigating the EBPT  in a QSM with U(1) symmetry, and then we verify the assumptions behind \eqref{eq:analytic_edge} by investigating numerically the many-body mobility edges in QSMs with and without the U(1) symmetry.

\paragraph{The Quantum Sun models.}
\label{sec:model}
The QSM considered in \cite{Suntajs22} and depicted in Fig.~\ref{fig:sketch}(b), consists of a set of ${1/2}$-spins  divided into two subsystems: $N$ spins in the delocalized core, i.e., in the ``sun'', enjoy all-to-all interactions, while $L$ other spins, forming rays of the sun,   do not interact among themselves. Each spin $j$ is coupled only to one randomly chosen spin $n_j$ of the sun. The j-th outer spin is positioned at distance $u_j$ from the sun, with $u_j$ chosen with a  uniform  probability from the interval $[j-\zeta,j+\zeta]$, whereas its coupling to the spin from the sun is set as ${g_0\alpha^{u_j}}$.
The outer spins  are also exposed to a random magnetic field chosen with a uniform distribution ${h_j\in[1-W,1+W]}$. Internal degrees of freedom of the sun are described by a ${2^N\times 2^N}$ GOE  matrix $R=\frac{\beta}{2}(B+B^T)$, where the matrix elements
$B_{ij}$ are sampled independently from the standard normal distribution. We follow the values of parameters chosen in \cite{Suntajs22} having: ${N=3,\zeta=0.2,g_0=1,\beta=0.3, W=0.5}$. The resulting Hamiltonian of the original QSM is:
\begin{equation}
\hat{H}=R\otimes\mathds{1}+\sum\limits_{j=1}^{L}g_0\alpha^{u_{j}}\hat{S}^{x}_{n_{j}} \hat{S}^{x}_{j}+\sum\limits_{j=1}^L h_j\hat{S}^z_j.
\label{eq:sun}
\end{equation}

We introduce the U(1) symmetric QSM that adds in a minimal way the conserved quantity  - the projection of total spin on the $z$ axis: ${\hat{S}^z_{\mathrm{tot}}=\sum_{j=1}^{L+N}\hat{S}_j^z}$. To this end two modifications of the original Hamiltonian, \eqref{eq:sun} are necessary. First, $\hat{S}^z_{\mathrm{tot}}$ conservation needs to be imposed on the couplings of the outer spins to the sun. To accomplish this, each $\hat S^x\hat S^x$ term is complemented with a $\hat S^y \hat S^y$ term with equal coupling, i.e.: ${\hat{S}^x_{n_j}\hat{S}^x_j\to\hat{S}^x_{n_j}\hat{S}^x_j+\hat{S}^y_{n_j}\hat{S}^y_j}$. Secondly, the sun, $R$, should also commute with $\hat{S}^z_{\mathrm{tot}}$.
This condition makes the matrix representing $R$ block 
diagonal, with blocks corresponding to eigenvalues of $\hat{S}^z_{\mathrm{tot}}$ and the block ranks equal to the multiplicity of these eigenvalues. The conservation is imposed by requiring that any matrix element outside those blocks vanishes, leading to $R_{\mathrm{cons}}$ matrix. The Hamiltonian of the model with conserved $\hat{S}^z_{\mathrm{tot}}$ reads explicitly:
\begin{equation}
\hat{H}_{\mathrm{cons}}=R_{\mathrm{cons}}\otimes\mathds{1}+\sum\limits_{j=1}^{L}g_0\alpha^{u_{j}}(\hat{S}^{x}_{n_{j}} \hat{S}^{x}_{j}+\hat{S}^{y}_{n_{j}} \hat{S}^{y}_{j})+\sum\limits_{j=1}^L h_j\hat{S}^z_j,\label{eq:cons_model}
\end{equation}
manifesting the U(1) symmetry. In the following, our 
analysis of the above model is restricted to the largest symmetry sector, i.e., ${S^z_{\mathrm{tot}}=0}$ for ${N+L}$ even and ${S^z_{\mathrm{tot}}=-1/2}$ for ${N+L}$ odd. 

\paragraph{Ergodicy breaking phase transition.}
To analyze the transition between ergodic and localized phases, we calculate eigenvalues $E_i$ using full exact diagonalization for the U(1) symmetric QSM \eqref{eq:cons_model}. Results are obtained using $10000,3000$ realizations of disorder for ${L=7-12,13-14}$, respectively. Additionally, $3000$ realizations were taken for $L=15-16$ using POLFED algorithm \cite{Sierant20p}. Note that the total length of the chain is $N+L$ reaching $19$ in our calculations.
From the spectrum we compute the gap ratios ${r_i=\min({s_i,s_{i+1}})/\max({s_i,s_{i+1}})}$ \cite{Oganesyan07, Atas13, Giraud22}, where ${s_i=E_{i+1}-E_i}$ is the level spacing. The gap ratios are averaged over ${n_{\mathrm{ev}}=}{\min(\dim \mathcal{H}/10,500)}$ eigenvalues from the middle of the spectrum, with $\dim\mathcal{H}$ being the dimension of the Hilbert space, and finally the average over all available disorder realizations is performed to obtain $\langle r \rangle$.
For ergodic systems, the value of this parameter is close to the GOE value $\langle r\rangle \approx 0.53$, while for localized systems $\langle r \rangle\approx 0.39$, as in the Poisson ensemble \cite{Atas13}. 

\begin{figure}[ht]
\centering
\includegraphics [width=0.45 \textwidth,angle=0]{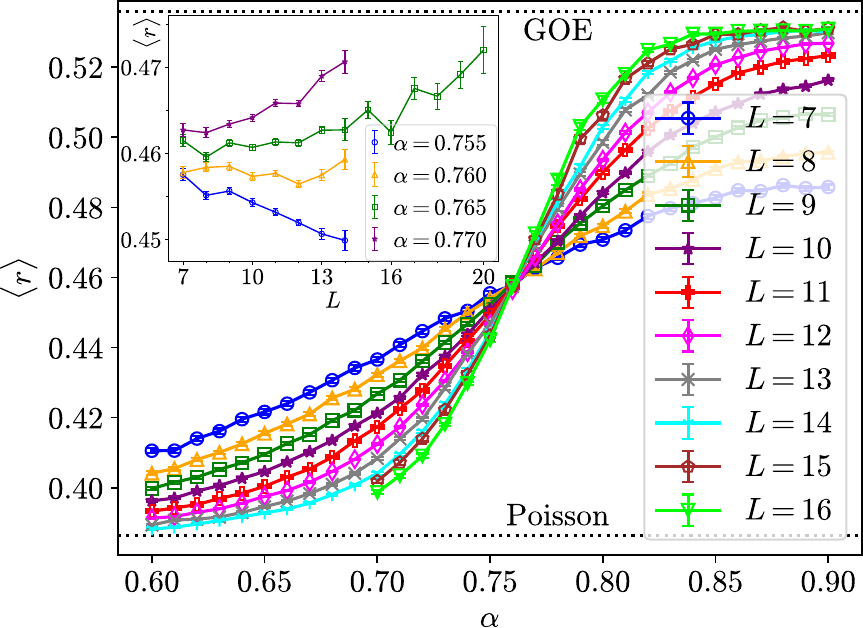}
\caption{Mean gap ratios $\langle r \rangle$  in the middle of the spectrum, for different couplings $\alpha$ and number of outside spins $L$ for Hamiltonian
\eqref{eq:cons_model}, cross practically in a single $L$ independent point at $\alpha_c\approx 0.76$. The inset shows the small variation of $\langle r \rangle$ with system size $L$ around critical $\alpha$ value. Black dotted lines indicate the values expected from ergodic (GOE) and integrable (Poisson) systems.} \label{fig:rbar}
\end{figure}

Our results are presented in Fig.~\ref{fig:rbar}. Conclusions of \cite{Suntajs22}  hold also for our  model \eqref{eq:cons_model}: there exists $\alpha_c\approx 0.76$ at which $\langle r \rangle$ is practically independent of the system size, and all $\langle r \rangle$ vs $\alpha$ curves intersect at this point. To identify the transition point $\alpha_c$ more precisely, we investigate $\langle r\rangle$ as a function of $L$ for fixed values of $\alpha$ close to $\alpha =0.76$, as shown 
in the inset of Fig.~\ref{fig:rbar}. We observe a clear monotonous dependence of $\langle r \rangle$ on $L$ for $\alpha=0.755$ and $\alpha=0.77$, which indicates that the system is, respectively, in the localized and ergodic regime at those values of $\alpha$. Data for $L\leq 14$ is inconclusive for $\alpha=0.765$. Hence,  using the POLFED algorithm~\cite{Sierant20p} we additionally calculate $500$ energies in the middle of the spectrum for ${L=15-19,20}$, using ${1000,250}$ disorder realizations, respectively, showing that the point $ \alpha = 0.765$ is in the ergodic phase. This allows us to narrow down the possible position of the critical point to ${\alpha_c\in(0.755,0.765)}$. 

The position of the critical point can be studied more quantitatively by extracting from Fig.~\ref{fig:rbar} the system size dependent 
intersection point, $\alpha_c(L)$, between $\langle r \rangle$ curves for system sizes $L-1$ and $L+1$. 
Such an analysis, see \cite{suppl}, reveals a slight shift of the intersection point with system size $L$, which is well described by $\alpha_c(L)=\alpha_c(\infty) + a/L$ with $\alpha_c(\infty)=0.749(7)$. This result is in agreement with the previously obtained bounds on $\alpha_c$. Similar mild system size dependence of the intersection point is also present in the original QSM~\cite{Suntajs22}. Even though system size dependence of the critical point is found, it is very subtle, in contrast to extended 1D systems studied in the context of MBL, discussed in the introduction. Moreover, we have verified that the same conclusions about ergodicity breaking in the  QSM \eqref{eq:cons_model} are obtained from analysis of the Thouless time extracted from the spectral form factor \cite{Haakebook,Suntajs20e,Sierant20b,Suntajs22}, see \cite{suppl}.

\paragraph{Numerical test of the mobility edge.}
\label{sec:ME}
The results obtained up till now clearly indicate the EBPT  in the thermodynamic limit at ${\alpha\approx 0.76}$ for the states in the middle of the spectrum of \eqref{eq:cons_model}. 
We now numerically test whether the QSMs exhibit a mobility edge, i.e. whether the transition point depends on the energy.
The QSMs, both in its original version \eqref{eq:sun} and in our,  the $\hat  S_z$ preserving model \eqref{eq:cons_model}, seem to be ideal candidates to study this problem due to the sharpness of the transition.

Having the extensive set of eigenvalues for different system sizes for  \eqref{eq:cons_model}, we diagonalized the QSM \eqref{eq:sun}  for $L=6-12$ and considered at least $3000$ disorder realizations. Then, we followed the procedure similar to the one described in \cite{Luitz15}. We determined the scaled eigenenergies,
${\varepsilon=(E-E_{\textrm{min}})/(E_{\textrm{max}}-E_{\textrm{min}})}$, where $E_{\textrm{min}}$ ($E_{\textrm{max}}$) are the smallest (largest) eigenvalue
for a particular diagonalization. We obtain mean gap ratios within  the energy windows centered at specified values of ${\varepsilon\in\{0.1,0.15,\ldots,0.9\}}$ instead of the center of the spectrum $\varepsilon=0.5$. For both models \eqref{eq:sun} and \eqref{eq:cons_model} we diagonalized two more system sizes using POLFED \cite{Sierant20p} and at least $3000$ disorder realizations.

\begin{figure}[ht]
\centering
\includegraphics [width=0.45 \textwidth,angle=0]{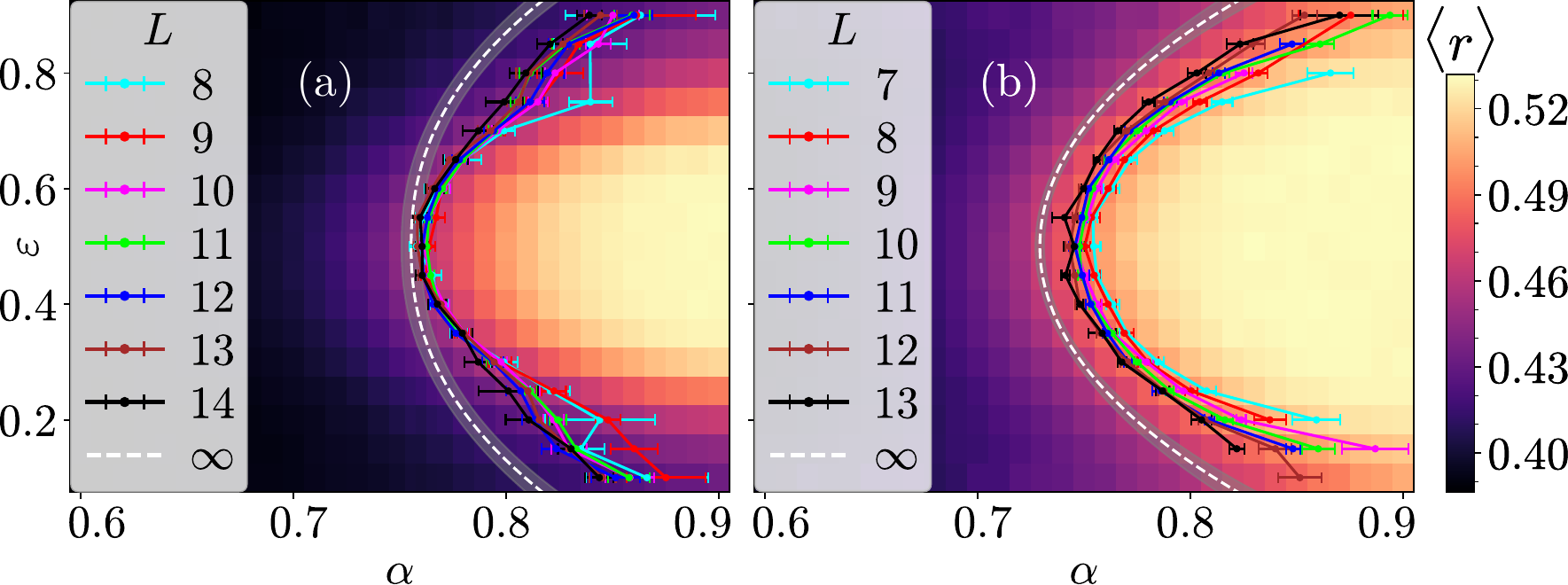}
\caption{Phase diagram for the QSM. (a) U(1)-preserving QSM \eqref{eq:cons_model}. (b) The original QSM \eqref{eq:sun}. Background colors show the value of $\langle r\rangle$ for biggest fully diagonalized system size considered for each model, namely $L=14$ for (a) and $L=12$ for the (b) (total system size is $L+3$ taking into account the GOE core). Solid lines show position of system size dependent boundary between phases $\alpha^{\varepsilon}_{c}(L)$ defined in the main text. Dashed white lines give $L\rightarrow\infty$ extrapolation of the border between localized (left, $\alpha<0.75$) and ergodic regimes (right), while the band around it corresponds to the $2\sigma$ confidence interval of the extrapolation, see \cite{suppl} for details.} \label{fig:ME_heatmap}
\end{figure}

\begin{figure}[ht]
\centering
\includegraphics [width=0.225 \textwidth,angle=0]{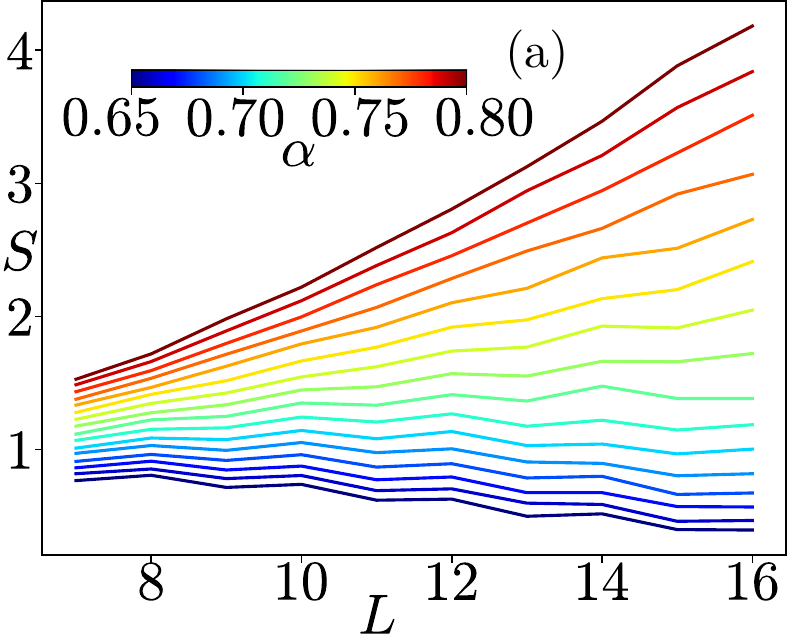}
\includegraphics [width=0.225 \textwidth,angle=0]{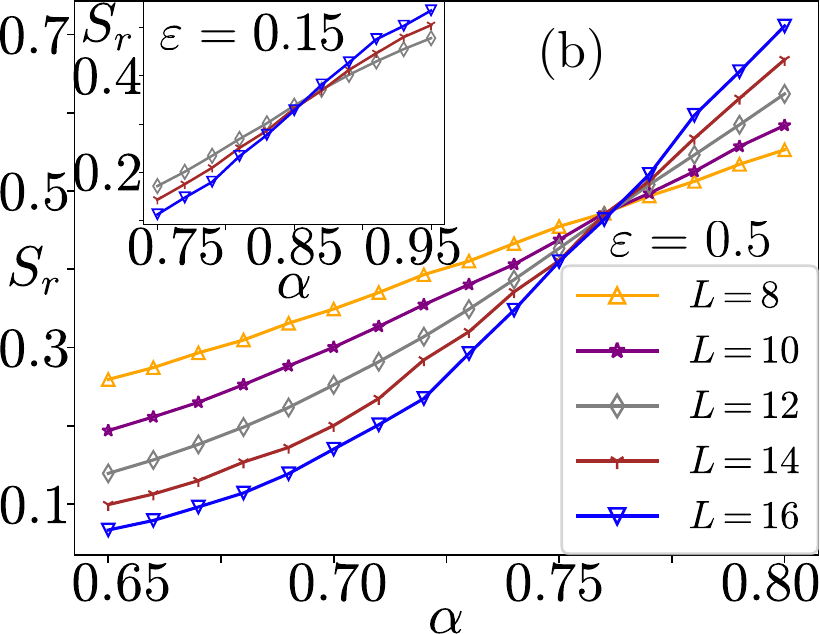}
\caption{Entanglement entropy of eigenstates in U(1) symmetric QSM \eqref{eq:cons_model}. (a) Average entropy, $S$, for the bipartition described in the main text, averaged over the middle of the spectrum (around $\varepsilon=0.5$), for different system
sizes $L$. Note the linear increase in the ergodic regime, flattening close to the transition at $\alpha\approx 0.76$ and even decreasing for smaller $\alpha$. 
(b) The rescaled entanglement entropies, $S_r$ (see text) for different system sizes cross at EBPT point for $\varepsilon= 0.5$. Inset presents the similar crossing for energies around $\varepsilon=0.15$ for $\alpha\approx 0.85$ indicating also the mobility edge.
}\label{fig:entanglement}
\end{figure}

The data obtained for different system sizes and  at different $\varepsilon$ are presented in \cite{suppl}. Here, we concentrate 
on  the thermodynamic limit only.
We follow the same procedure as for $\varepsilon=0.5$ above, namely, we extract the crossing point $\alpha_c^\varepsilon(L)$ between the average gap ratios for systems of size $L-1$ and $L+1$, at given $\varepsilon$. More consistent results were obtained when only a fraction of the energy window size was considered, namely $0.25n_{\mathrm{ev}}$ for the original \eqref{eq:sun} and $0.5n_{\mathrm{ev}}$ for the U(1) symmetric \eqref{eq:cons_model} models. Results presented in Fig.~\ref{fig:ME_heatmap} show that the further from the middle of the spectrum, the bigger the value of $\alpha_c^\varepsilon(L)$.  
Moreover, with increasing system size, the critical point for all values of $\varepsilon$ noticeably shifts towards smaller values of $\alpha$.  To extrapolate the results to the thermodynamic limit in the QSM \eqref{eq:sun}, we use a two parameter formula $\alpha_c(\varepsilon)= A_0 e^{\frac{( \varepsilon-1/2)^2}{A_1}}$ inspired by \eqref{eq:analytic_edge}. We find that the two-parameter formula matches the numerical data perfectly at any system size. 
This indicates that the mobility edge in that version of the model is caused solely by the variation of the density of states with $\varepsilon$.  Finally, we extract the fit parameters for each system size and extrapolate them to the thermodynamic limit leading to the mobility edge shape presented in Fig.~\ref{fig:ME_heatmap}(b). The same reasoning, after slight modification of the assumed density of states \cite{suppl}, applies to the U(1) symmetric QSM \eqref{eq:cons_model}, with the result shown in Fig.~\ref{fig:ME_heatmap}(a).

\paragraph{Entanglement entropy of eigenstates.}
EBPT should manifest itself also in the properties of eigenvectors. We concentrate here on the entanglement entropy of eigenstates obtained by splitting the system in two parts $A$ and $B$ 
and finding $S(|\psi\rangle) = - \Tr_A \rho_A \ln \rho_A$ where $\rho_A=\Tr_B |\psi\rangle\langle \psi|$.  A small subtlety arises as the parts cannot be equal, one of them contains the GOE sun. We consider the splitting in the middle of the system, with the part $A$ consisting of the GOE sun and $\lfloor(N+L)/2\rfloor-N$ spins closest to it, and the rest of the spins in the part $B$, here $\lfloor \cdot \rfloor$ denotes the floor function. The dependence of average value of entropy, $S$, for the middle part of the spectrum, as a function of the system size is shown in Fig.~\ref{fig:entanglement} for the $U(1)$ preserving model. In the extraction of the average value around $\varepsilon$, states lying in the rescaled energy interval $[\varepsilon-0.025,\varepsilon+0.025]$ were considered only. Note the visible even-odd oscillations resulting from taking different $S^z_{\rm tot}$ sectors for even and odd
$L$. We observe that $S$ for $\alpha> \alpha_c$ grows almost linearly with the system size, following a volume law. In contrast, $S(L)$ becomes practically flat below $\alpha_c$
(area law behavior) even showing a decrease with the system size. The latter effect is due to the fact that with increasing $L$, the splitting between $A$ and $B$ performed in the middle of the system moves further from the GOE sun, which is the source of entanglement growth in the system.

Scaling the entanglement entropy with the Page value $S_P$ \cite{Page93, bianchi_hackl_22} i.e. the typical value for a random vector from the Hilbert space, we obtained the rescaled entanglement entropy, $S_r=S/S_P$ c.f. Fig.~\ref{fig:entanglement}(b). For 
$S_r$ corresponding to the middle of the spectrum the curves for different system sizes cross around $\alpha_c$ in a good correspondence with gap ratio statistics.
The inset shows a similar crossing for states around $\varepsilon\approx 0.15$ which is significantly shifted to larger values of $\alpha$ agreeing with the data of Fig.~\ref{fig:ME_heatmap}, providing another evidence for the existence of the mobility edge. Similar data is obtained for the original QSM \cite{suppl}.

\paragraph{Conclusions.}
The aims of the present work are twofold. First,
we have shown that it is possible to construct the QSM with a conserved total spin projection on the $z$ axis. Translating the model from spin 
to particle language (via e.g. Jordan-Wigner transformation)
results in the model with particle number conservation, more prone to experimental realizations.  We have shown that the resulting U(1) symmetric QSM shares with its predecessor all the essential features resulting in the clear, unambiguous transition between ergodic and localized phases. In particular, the introduced model allowed us to demonstrate that the U(1) symmetry affects the EBPT transition in QSM in a minor way.
The second aim of  this work 
is to provide the evidence for the existence of the many-body mobility edge in the thermodynamic limit. This is indeed confirmed for both considered QSMs by the extrapolation of our numerical results, aided by analytic formulas derived from the hybridization condition for QSM.

Over the last decades, the mobility edges in non-interacting quantum systems have been a focus of theoretical studies~\cite{Sarma88, Biddle10, Biddle11, Delande14, Pasek17, Li17, Li20} and experimental investigations~\cite{Kondov11Three, Jendrzejewski12, Semeghini15, Luschen18, Kohlert19}. Our work, employs the lack of finite size drifts in the quantum sun models, to show an unambiguous evidence for many-body mobility-edge in a strongly interacting system. 
The discussed mechanism for appearance of the stable many-body mobility edge relies on the exponential suppression of many-body density of states away from the middle of the spectrum. A question of whether this mechanism could lead to stable many-body mobility edges in other thermodynamically large disordered many-body systems remains an outstanding challenge for future investigations.

\paragraph{Note added.} Since the original submission of this Letter a related {work} appeared \cite{Suntajs24} which revealed the similarity between the original QSM and the ultrametric ensemble. It finds multifractality of eigenstates at the transition (consistent with the analysis in~\cite{hopjan_23}) and Fock space localization of eigenstates beyond EBPT.  {On the other hand, it has been {argued} \cite{Colbois24} that additional insight into the MBL transition may be obtained from density-density correlation functions. It will be also a promising direction for future studies {of the QSM}.

\acknowledgements
We gratefully acknowledge discussions with W.~De Roeck and A.S. Aramthottil.
We also acknowledge Poland’s high-performance computing infrastructure PLGrid (HPC Centers: ACK Cyfronet AGH) for providing computer facilities and support within computational grant no. PLG/2023/016370. This research was
 funded by National Science Centre (Poland) under the OPUS call within the WEAVE programme
2021/43/I/ST3/01142 (K.P. and J.Z.). For the purpose of
Open Access, the authors applied a CC-BY public copyright
licence to any Author Accepted Manuscript (AAM) version arising from this submission.
A partial support by the Strategic Programme Excellence
Initiative at Jagiellonian University is noted.
P.S. acknowledges support
from ERC AdG NOQIA; MICIN/AEI (PGC2018-0910.13039/501100011033, CEX2019-000910-S/10.13039/501100011033, Plan National FIDEUA PID2019-106901GB-I00, FPI; MICIIN with funding from European Union NextGenerationEU (PRTR-C17.I1): QUANTERA MAQS PCI2019-111828-2); MCIN/AEI/ 10.13039/501100011033 and by the “European Union NextGeneration EU/PRTR" QUANTERA DYNAMITE PCI2022-132919 within the QuantERA II Programme that has received funding from the European Union’s Horizon 2020 research and innovation programme under Grant Agreement No 101017733Proyectos de I+D+I “Retos Colaboración” QUSPIN RTC2019-007196-7); Fundació Cellex; Fundació Mir-Puig; Generalitat de Catalunya (European Social Fund FEDER and CERCA program, AGAUR Grant No. 2021 SGR 01452, QuantumCAT \ U16-011424, co-funded by ERDF Operational Program of Catalonia 2014-2020); Barcelona Supercomputing Center MareNostrum (RES-FI-2024-1-0043); EU (PASQuanS2.1, 101113690); EU Horizon 2020 FET-OPEN OPTOlogic (Grant No 899794); EU Horizon Europe Program (Grant Agreement 101080086 — NeQST), ICFO Internal “QuantumGaudi” project; European Union’s Horizon 2020 research and innovation program under the Marie-Skłodowska-Curie grant agreement No 101029393 (STREDCH) and No 847648 (“La Caixa” Junior Leaders fellowships ID100010434: LCF/BQ/PI19/11690013, LCF/BQ/PI20/11760031, LCF/BQ/PR20/11770012, LCF/BQ/PR21/11840013).
L.V. acknowledges support by the Slovenian Research and Innovation Agency (ARIS), Research core funding Grants No.~P1-0044, N1-0273 and J1-50005.
Views and opinions expressed in this work are, however, those of the authors only and do not necessarily reflect those of the European Union, European Climate, Infrastructure and Environment Executive Agency (CINEA), nor any other granting authority. Neither the European Union nor any granting authority can be held responsible for them. No part of this work was written with the help of artificial intelligence software.

%

\newpage

\newcommand{\snum}{S}

\renewcommand{\theequation}{\snum.\arabic{equation}}
\renewcommand{\thefigure}{\snum.\arabic{figure}}

\setcounter{equation}{0}
\setcounter{figure}{0}

 \pagebreak
 \section{Supplementary material for Many-body mobility edge in Quantum Sun models}
 \label{appendix1}

\section{Intermediate statistics}
Both average gap $\langle r \rangle$ ratio and entanglement entropy $S$ of eigenstates in the QSM reveal, that in the transition point between ergodic and localized regimes almost no system size dependence can be seen. On the other hand, more detailed analysis of the spectral statistics shows that probability distributions of both level spacing $s$ and gap ratio $r$ approach respectively Poissonian prediction in the localized phase, and GOE prediction in an ergodic one. At a finite system size those distributions are in fact intermediate between both model cases of GOE and Poissonian statistics, thus it constitutes a valid question what will be the fate of those probability distributions in the critical point once the thermodynamic limit is approached. 

To investigate this question, we aim to calculate numerical distributions of the spacing ratio $P(s)$, although doing so requires first getting rid of the influence of the DoS present in the spectrum. After obtaining an ordered set of eigenvalues $\{E_n\}$, this can be achieved through so-called unfolding of the spectrum procedure and an ordered set of unfolded eigenvalues $\{\widetilde E_n\}$ is obtained. Unfolding is done by calculating the cumulative density of states function ${G(E)=\sum_n\Theta(E-E_n)}$, where $\Theta$ is the Heaviside step function, and then separating it into continuous and fluctuating part ${G(E)=\overline{G}(E)+\delta G(E)}$, where $\overline{G}(E)$ is taken as a polynomial of small degree $n_p$ fitted to $G(E)$. Degree $n_p$ is chosen as the smallest one that gives average spacing $\approx 1$ after the unfolding, which corresponds in all considered cases to $n_p=3,4$. Unfolded eigenvalues are defined as $\widetilde E_n=\overline{G}(E_n)$. In principle, the continuous part $\overline{G}$ might exhibit nonphysical regions, where density of states would be decreasing. To get rid of this numerical artifact $\widetilde E_n$ is included in calculation of SFF only if, after adding it, the set $\{\widetilde E_n\}$ is still ordered in nondecreasing manner.

After unfolding, spacing $s$ is calculated for $n_{\rm ev}=\min\{500,\dim\mathcal{H}/10\}$ energies from the middle of the spectrum to give a numerical distribution $P(s)$, shown for the critical value of $\alpha$ in Fig.~\ref{fig:intermediate}. Numerical distributions were fitted with {$\nu-h$ distribution \cite{Sierant20} (originally called $\beta-h$, here renamed to avoid confusion with parameter $\beta$ of the main text), interpolating between GOE $(\nu=1,h\to\infty)$ and Poisson $(\nu=0)$ statistics, recently shown to well describe the EBPT. }
\begin{figure}[ht!]
\centering
\includegraphics [width=0.225 \textwidth,angle=0]{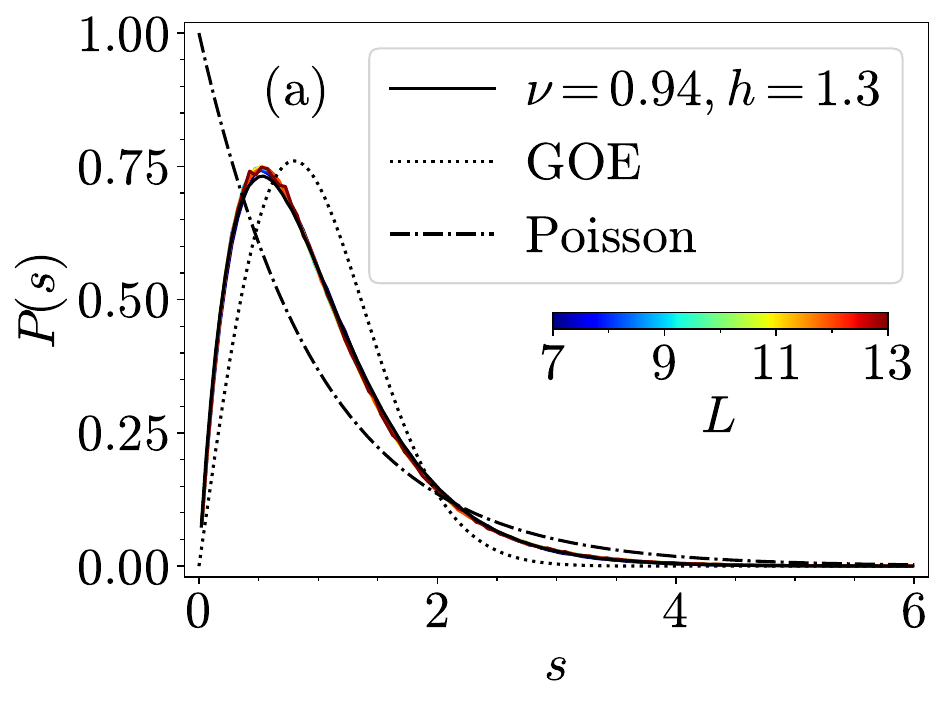}
\includegraphics [width=0.225 \textwidth,angle=0]{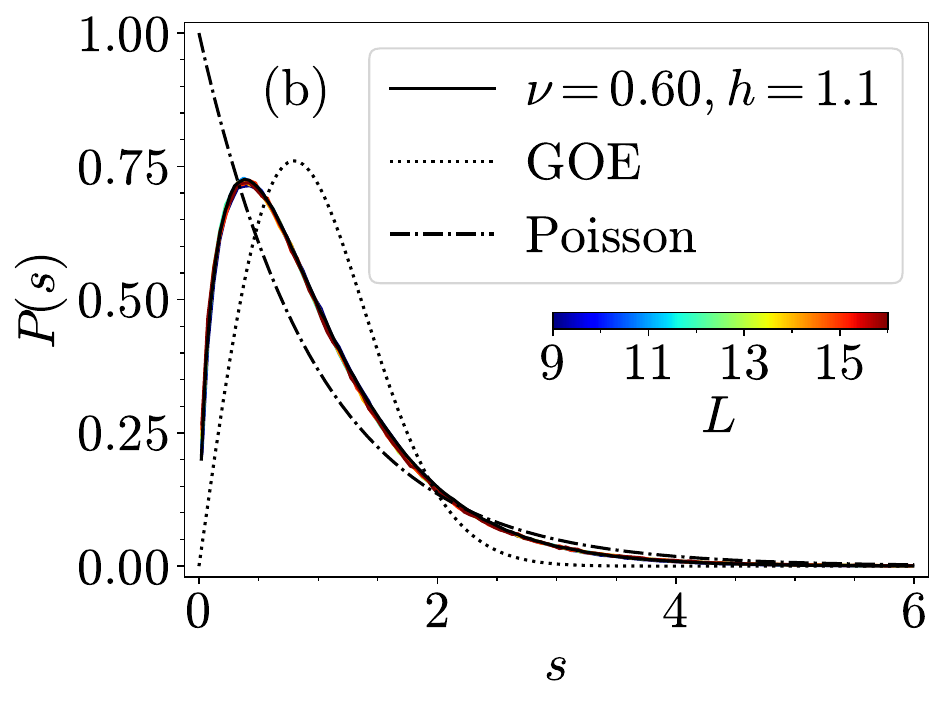}
\caption{Intermediate level statistics close to the transition point. Colored lines show numerical distributions. (a)~Original quantum sun model at $\alpha=0.75$. (b)~U(1) symmetric model at $\alpha=0.76$. Black lines show GOE, Poisson and $\nu-h$ distribution (see text), the latter of which was fitted to the numerical data, giving $\nu$ and $h$ shown in the corresponding legends.}\label{fig:intermediate}
\end{figure}
For both studied variants of the QSM, the level spacing distribution did not show any signs of system size dependence at the critical point, and excellent agreement with the theoretical {$\nu-h$ distribution was observed. Moreover, $0<\nu<1$  in both models proves that critical statistics is neither of chaotic nor integrable system. Investigation of the tails of numerical distributions reveals that they perfectly follow the $\nu-h$ fit.}

\section{Spectral form factor}

To further confirm presence of the localization transition, we use a different indicator of ergodicity breaking, the spectral form factor (SFF) \cite{Haakebook,Suntajs20e,Sierant20b,Suntajs22}, defined as:
\begin{equation}
K(\tau)=\frac{1}{Z} \left\langle\left|
\sum\limits_{n=1}^{\dim\mathcal{H}} \rho(\widetilde E_n)e^{-i2\pi\widetilde E_n\tau}
\right|^2\right\rangle,\label{eq:SFF}
\end{equation}
where $\widetilde E_n$ are eigenvalues after the unfolding procedure \cite{Gomez02}, $\rho(\widetilde E)$ is a Gaussian weight diminishing the importance of spectral edges, $Z$ is a normalization constant, such that ${K(\tau\gg 1) \approx 1}$ and the average is performed over disorder realizations. The SFF is known analytically for GOE \cite{Mehtabook} as ${K_{\mathrm{GOE}}(\tau)=2\tau-\tau\ln(1+2\tau)}$ for ${\tau\leq1}$ and ${K_{\mathrm{GOE}}(\tau)=2-\tau\ln[(2\tau+1)/(2\tau-1)]}$ for ${\tau>1}$. For many-body systems enjoying time reversal symmetry there exists a time $\tau_{\mathrm{Th}}$, called a Thouless time, beyond which the SFF follows the GOE prediction. It is worth stressing, that due to the unfolding procedure, $\tau$ present in the definition \eqref{eq:SFF} is in fact dimensionless and related to the real physical time via the Heisenberg time, $t_H$ (the inverse of the mean level spacing)  as $\tau=t/t_H$. 

In calculation of SFF, procedure of \cite{Suntajs22} is followed. After unfolding the spectrum, SFF \eqref{eq:SFF} is calculated using a spectral filter diminishing  the influence of spectral edges $\rho(\widetilde E)=\exp[{-(\widetilde E- \langle\widetilde E\rangle)^2/(2\eta^2\sigma_{\widetilde E}^2)}]$, with $\langle \widetilde E\rangle$ and $\sigma_{\widetilde E}^2$ being mean and variance in a given realization respectively. Parameter $\eta$ controls effective fraction of included eigenvalues, $\eta=0.5$ is taken. To carry out the sum in \eqref{eq:SFF}, a nonuniform fast Fourier transform algorithm is used. Normalization $K(\tau\gg 1)\approx 1$ in \eqref{eq:SFF} is assured by putting $Z=\left\langle\sum_{n=1}^{\dim \mathcal{H}} |\rho(\widetilde E_n)|^2\right\rangle$.

\begin{figure}[ht]
\centering
\includegraphics [width=0.225 \textwidth,angle=0]{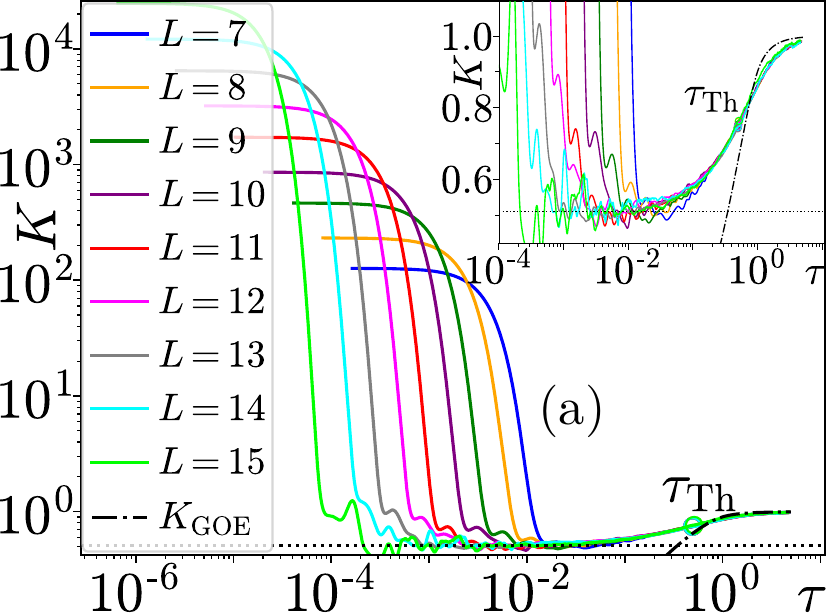}
\includegraphics [width=0.225 \textwidth,angle=0]{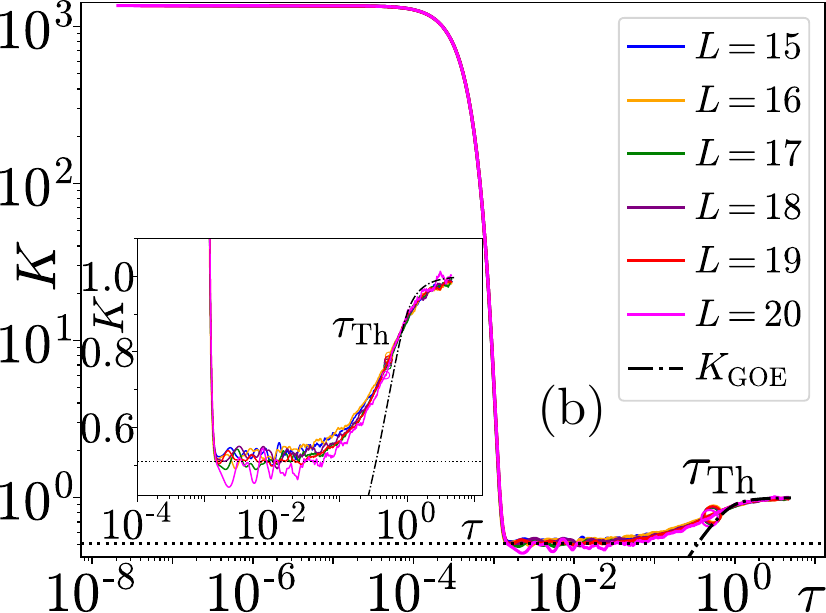}
    \caption{SFF in the U(1)-preserving QSM (5) for ${\alpha=0.765}$. Small sizes data (a) are from full exact diagonalization, larger $L$ data (b) obtained using  POLFED algorithm \cite{Sierant20p} finding $2500$ eigenvalues in the middle of the spectrum. Insets show zooms on corresponding main panels. The horizontal dotted line corresponds to the value in the constant regime ${K(\tau)\approx 0.51}$. Circles indicate the obtained values of Thouless times. }\label{fig:cons_sff}
\end{figure}

Following \cite{Suntajs22}, $K(\tau)$ was calculated for $5000$ times $\tau_i\in[1/(2\pi\dim\mathcal{H}),5]$, equidistant in logarithmic scale. As shown in the literature \cite{Cotler17}, the SFF does not have self-averaging property, so it has fluctuations that do not decrease with increasing number of realizations. Because of that, an additional moving time average is performed to obtain $\overline{K}(\tau_i)=\sum_{j=i-w}^{i+w}K(\tau_j)$ for $w=25$, which reduces the list of $\tau_i$ to $4950$ points. The difference from GOE is given by:

\begin{equation}
\Delta K(\tau)=\left|
\ln\left(
\frac{
\overline{K}(\tau)}
{K_{ \mathrm{GOE}}(\tau) }
\right)
\right|.
\end{equation}
The Thouless time $\tau_{\mathrm{Th}}$ is defined to be the biggest time for which $\Delta K(\tau)\geq \epsilon$, where, $\epsilon=0.06$ was chosen based on the criterion that $\Delta K(\tau_{\mathrm{Th}})$ must be larger than the noise.

The analysis of the SFF shape (Fig.~\ref{fig:cons_sff}) for a value close to the critical region ${\alpha=0.765}$ reveals that the SFF does not exhibit dependence on system 
size. Here, apart from exact diagonalization, we use the POLFED algorithm \cite{Sierant20p} to reach larger system sizes up to $L=20$. In the latter case, we extract $2500$ eigenvalues in the middle of the spectrum, with ${1000,250}$ realizations for ${L=15-19,20}$ respectively. This does not allow determining $K(\tau)$ for small $\tau$, as the regime before the initial drop (e.g. $\tau<10^{-3}$ on right panel of Fig.~\ref{fig:cons_sff}) carries no physical information. However, this is not an interesting range of $\tau$ values. Right after the initial drop $K(\tau)$ is constant and $L$ independent over a range of $\tau$ values.

The behavior of SFF observed closely resembles the one reported for an original quantum sun model \cite{Suntajs22} showing, without any doubt, that the conservation of global $S_z$ projection in the model (5) does not affect the character of the ergodicity breaking transition.

In case of larger systems $L=16-20$  instead of the whole spectrum only $n_e=2500$ eigenvalues from the middle of the spectrum are included in \eqref{eq:SFF}, which in general for $n_e\ll\dim\mathcal{H}$ has the effect of shifting the initial drop of SFF to bigger time $\tau_{n_e}$ \cite{Sierant20b}, but for $\tau\gg \tau_{n_e}$, the SFF is correctly reproduced. As seen in Fig.~\ref{fig:cons_sff}, $\tau_{n_e=2500}\approx 10^{-3}$, and since the region interesting for extraction of Thouless time is $\tau>10^{-1}$, so the obtained value of $\tau_{\mathrm{Th}}$ is not affected by limited number of eigenvalues.

\begin{figure}[ht]
\centering
\includegraphics [width=0.45 \textwidth,angle=0]{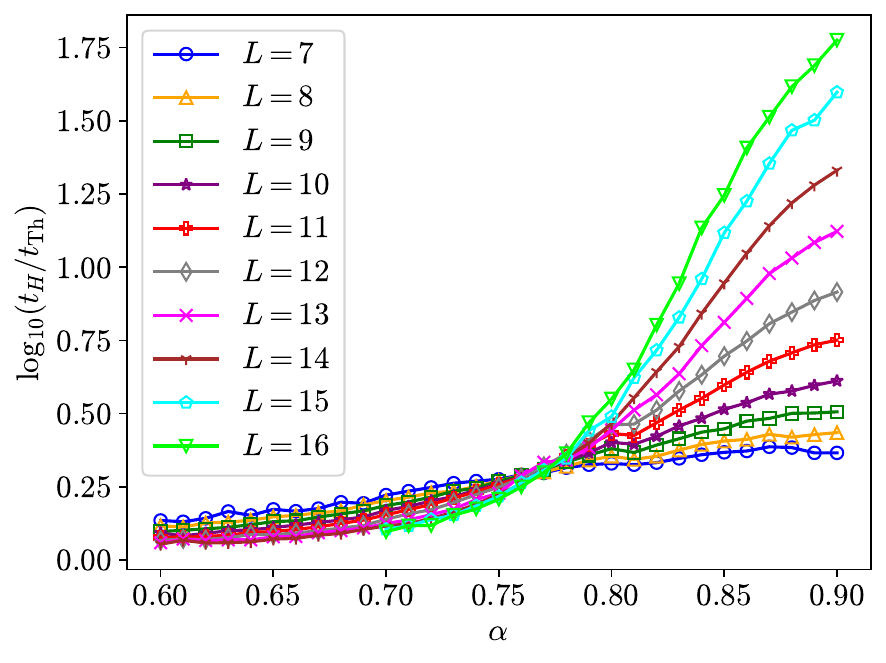}
\caption{Values of $t_H/t_{\mathrm{Th}}=1/\tau_{\mathrm{Th}}$ for different couplings $\alpha$ and number of outside spins $L$ for the U(1)-preserving QSM (5).}\label{fig:tau_th}
\end{figure}

The numerically obtained SFF easily allows for extracting the Thouless times for all analyzed values of $\alpha\in \{0.6,0.61,\ldots,0.9\}$. 
 The results for different system sizes are presented in Fig.~\ref{fig:tau_th}. The data cross almost at a single, size independent point ${\alpha\approx 0.76}$, in an excellent agreement with the gap ratio analysis.

\section{Calculation of statistical uncertainties}

In our consideration of the MBL phase transition in the main text, we calculate the gap ratios $r$ and average them over a small energy window around the rescaled energy $\varepsilon$, obtaining a single realization average $r_S$. It is known that single realization average is not self-averaging quantity \cite{TorresHerrera20,TorresHerrera20b}, meaning that the variance $\langle (r_S-\langle r_S\rangle)^2\rangle$, where $\langle\cdot\rangle$ denotes disorder average, is not necessarily decreasing when system size is increased. Instead, disorder average $\langle r_S\rangle$ is considered, denoted $\langle r \rangle$ for short. Assuming that $r_S$ for different realizations are uncorrelated, the standard deviation of $\langle r \rangle$ reads \cite{Sierant23f}:

\begin{equation}
    \sigma_r= \frac{\langle (r_S-\langle r \rangle)^2\rangle^{1/2}}{N_{\mathrm{dis}}^{1/2}},
\end{equation}
where $N_{\mathrm{dis}}$ is the number of disorder realizations used in calculation of $\langle r\rangle$.

\begin{figure}[ht]
\centering
\includegraphics [width=0.45 \textwidth,angle=0]{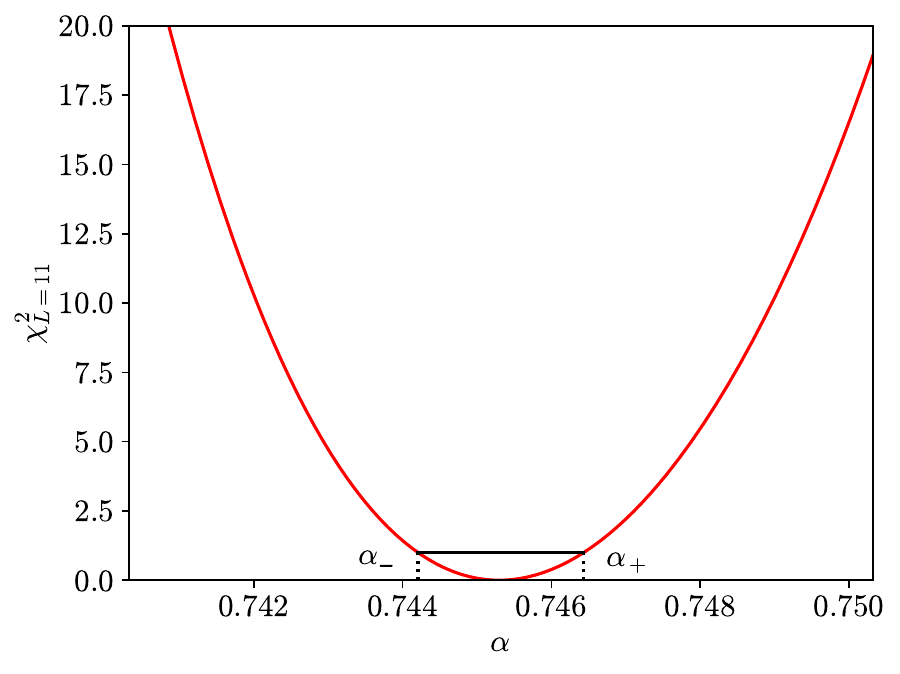}
\caption{Function $\chi^2_{L=11}$ in the vicinity of the crossing point of $\langle r \rangle$ vs $\alpha$ curves at $\varepsilon=0.5$ for the system sizes $L=10$ and $L=12$ in the original QSM (4). Horizontal line segment shows $\chi^2=1$ and positions of roots of this equation $\alpha_\pm$ are shown with dotted lines.} \label{fig:chi}
\end{figure}

Having uncertainties of data points on $\langle r \rangle$ vs $\alpha$ curves, we can estimate statistical errors for $\alpha_c^\varepsilon(L)$, which requires determining the crossing point of two polynomials $f_{L-1}(\alpha)$ and $f_{L+1}(\alpha)$, which approximate $\langle r \rangle$ vs $\alpha$ curves close to the critical point for $L-1$ and $L+1$, respectively. Using standard error propagation, we calculate errors of $f_{L-1},f_{L+1}$ as a function of $\alpha$, obtaining $\sigma_{L-1}(\alpha)$ and $\sigma_{L+1}(\alpha)$ correspondingly. Based on those functions the $\chi^2_L$ function is defined:
\begin{equation}
    \chi^2_L(\alpha)=\frac{[f_{L-1}(\alpha)-f_{L+1}(\alpha)]^2}{\sigma_{L-1}^2(\alpha)+\sigma_{L+1}^2(\alpha)}.
\end{equation}
Exactly at the intersection point $\chi^2_L=0$, and it increases approximately quadratically with the distance from minimum (see Fig.~\ref{fig:chi}), thus the equation $\chi^2_L(\alpha)=1$ has two solutions $\alpha_-<\alpha_+$. The quantity $(\alpha_+-\alpha_-)/2$ has statistical interpretation \cite{Cowan98} as the standard deviation of the position of the crossing point $\alpha_c^\varepsilon(L)$. All other instances of uncertainty calculation follow simply standard uncertainty propagation.

\section{Extraction and scaling of critical point}

Considering a plot of mean gap ratios $\langle r \rangle$ around the middle of the spectrum in the QSM (see Fig.~2 in the main text) it seems that all curves intersect at a single point, thus suggesting total independence of system size at the critical point between the ergodic and MBL phase. Instead of the middle of the spectrum, gap ratio can be calculated around rescaled energy $\varepsilon\in[0,1]$ (introduced in the main text), results are summarized in Fig.~\ref{fig:mobility_edge}.

\begin{figure}[ht]
\centering
\includegraphics [width=0.45 \textwidth,angle=0]{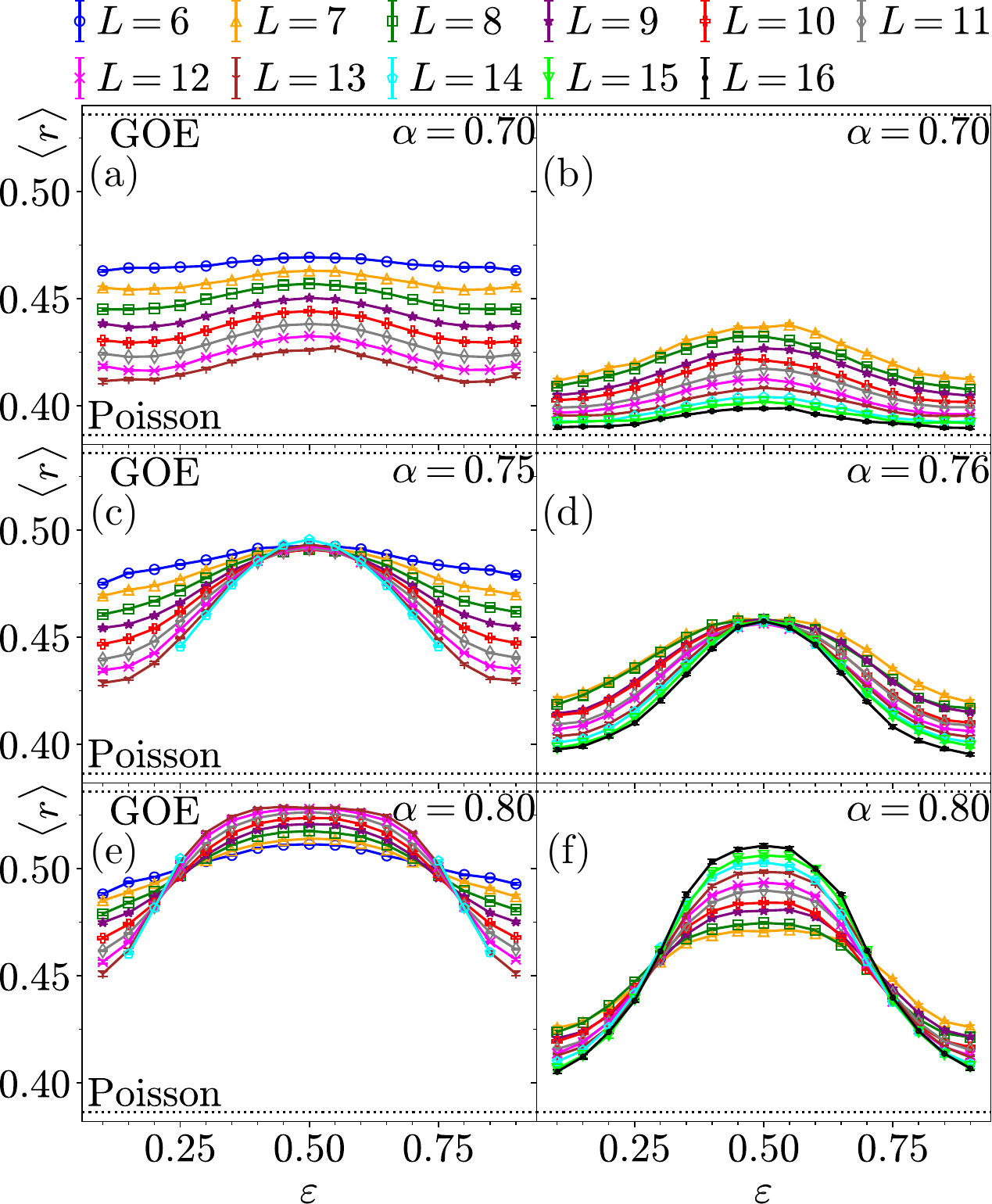}
\caption{Average gap ratio $\langle r \rangle$ as a function of position in the spectrum $\varepsilon$. Left column (a,c,e): the original QSM (4). Right column (b,d,f): the U(1)-preserving QSM (5). 
Upper panels (a,b) show localized regime, panels (c,d) show data in the vicinity of the critical point while panels (e,f) have the ergodic regime at the center and the localized regions at the edges.
Black dotted lines indicate limiting values for ergodic (GOE) and localized (Poisson) systems.}
\label{fig:mobility_edge}
\end{figure}

The right column, panels (b,d,f), in Fig.~\ref{fig:mobility_edge} shows data for the U(1) symmetric QSM (5). For $\alpha=0.7<\alpha_c$, all states are localized, as shown in Fig~\ref{fig:mobility_edge}(b).  In that case, with increasing system size, the mean gap ratio decreases towards the Poisson limit regardless of the energy. In Fig.~\ref{fig:mobility_edge}(d), the results at the critical point ${\alpha_c=0.76}$ are shown. Indeed, we observe that the mean gap ratio exhibits almost no size dependence for ${\varepsilon=0.5}$, i.e., at the center of the spectrum, while the rest of the spectrum shows tendency towards localization.

The situation is, however, manifestly different in Fig.~\ref{fig:mobility_edge}(f), for ${\alpha=0.8>\alpha_c}$. Here, at the center of the spectrum, the larger the system size, the bigger the $\langle r \rangle$. This clearly indicates that the system becomes ergodic with increasing $L$. On the other hand, for either low ${\varepsilon <0.3}$ or large ${\varepsilon >0.7}$ energies the trend is reversed, the bigger the system, the closer level statistics is to the many-body localized, Poisson limit. This is a clear manifestation of the many-body mobility edge in the system. 
As shown in the left column, panels (a,c,e), of Fig.~\ref{fig:mobility_edge}
the situation is  similar for the original QSM, (4), with all states being localized for small $\alpha$ [cf.~Fig.~\ref{fig:mobility_edge}(a)], no size dependence for $\alpha=\alpha_c$  at $\varepsilon=0.5$ [cf.~Fig.~\ref{fig:mobility_edge}(c)], and a clear indication of the mobility edge for $\alpha> \alpha_c$ [cf.~Fig.~\ref{fig:mobility_edge}(e)].

\begin{figure}[ht!]
\centering
\includegraphics [width=0.45 \textwidth,angle=0]{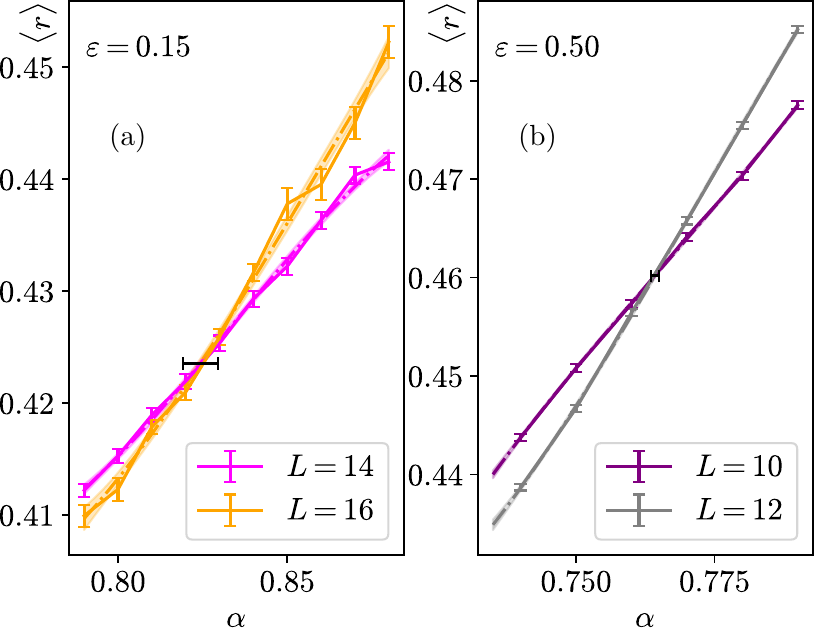}
\caption{Extraction of intersection points for the U(1)-preserving QSM (5) close to the edge of the spectrum (a) and the middle (b). The dash-dotted line is interpolation of data points with polynomial of degree $3$ and the shaded region corresponds to the $1\sigma$ confidence interval. The black horizontal line segment corresponds to the position and uncertainty of the crossing point.} \label{fig:intersections}
\end{figure}

An investigation of $\langle r \rangle$ vs $\alpha$ curves for different values of $\varepsilon$ yields, qualitatively, the same conclusions as for the center of the spectrum. To see whether those preliminary conclusions are correct, a more systematic approach needs to be applied. To that end, we extract positions of intersections of $\langle r \rangle$ vs $\alpha$ curves for different $L$, namely the intersection between $L-1$ and $L+1$ curves for a given $\varepsilon$, and we define position of this crossing to be system size and energy dependent critical point $\alpha_c^\varepsilon(L)$. Position of the intersection is obtained by fitting a polynomial of degree $3$ to the corresponding $\langle r \rangle$ curves in the interval $\alpha\in[\alpha_*^\varepsilon-0.05,\alpha_*^\varepsilon+0.05]$, where $\alpha_*^\varepsilon$ is the estimate of the intersection point for a given $\varepsilon$. Then $\alpha_c^\varepsilon(L)$ is extracted as the intersection point of fitted polynomials, see Fig.~\ref{fig:intersections}.

\begin{figure}[ht!]
\centering
\includegraphics [width=0.225 \textwidth,angle=0]{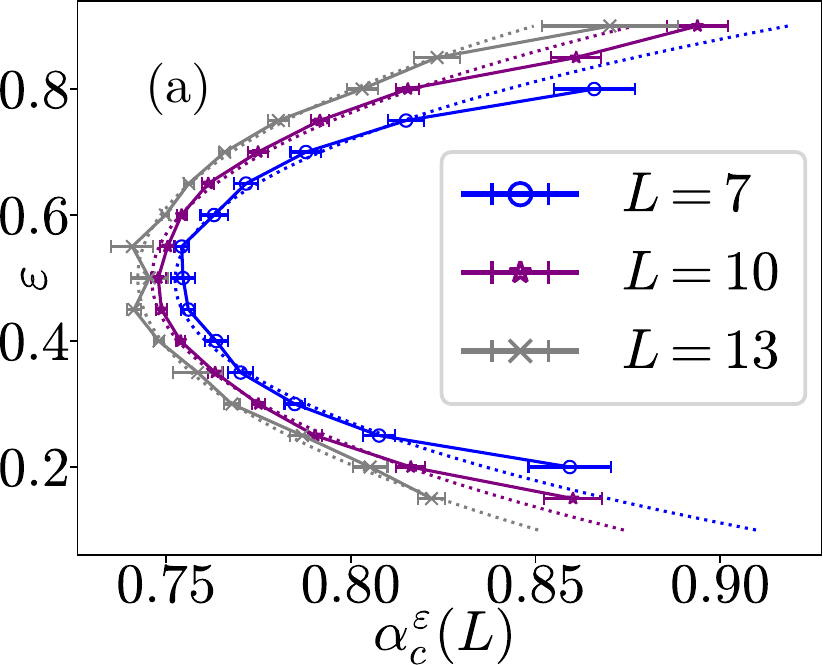}
\includegraphics [width=0.225 \textwidth,angle=0]{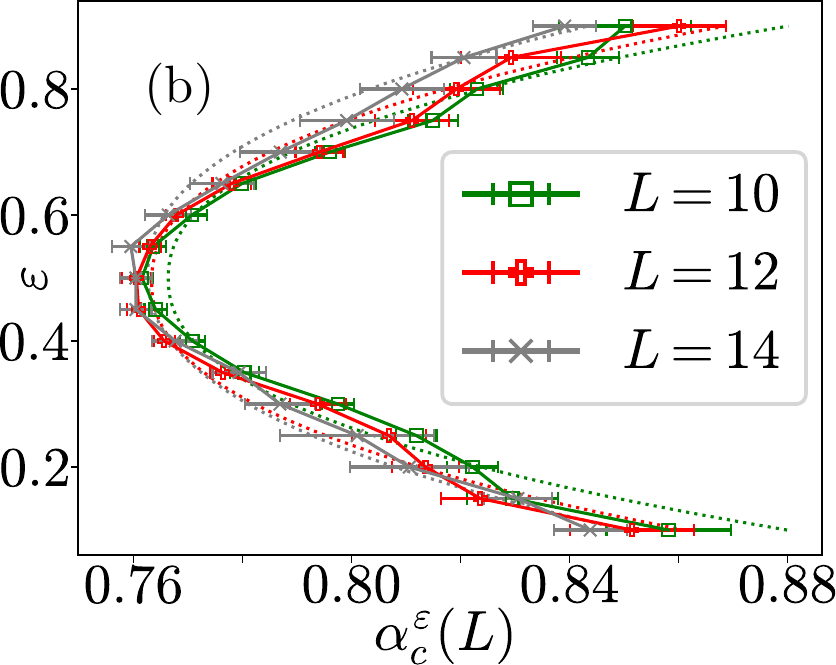}
\caption{Boundary between the phases in the QSM. (a) The original QSM (4). (b) The U(1)-preserving QSM (5). Transition points extracted from the data are shown with solid lines. Dotted lines show the fitted dependence on $\varepsilon$, according to Eq.~\eqref{eq:ME_analytical_fit}.}\label{fig:analytic_fit}
\end{figure}

\begin{figure}[ht]
\centering
\includegraphics [width=0.225 \textwidth,angle=0]{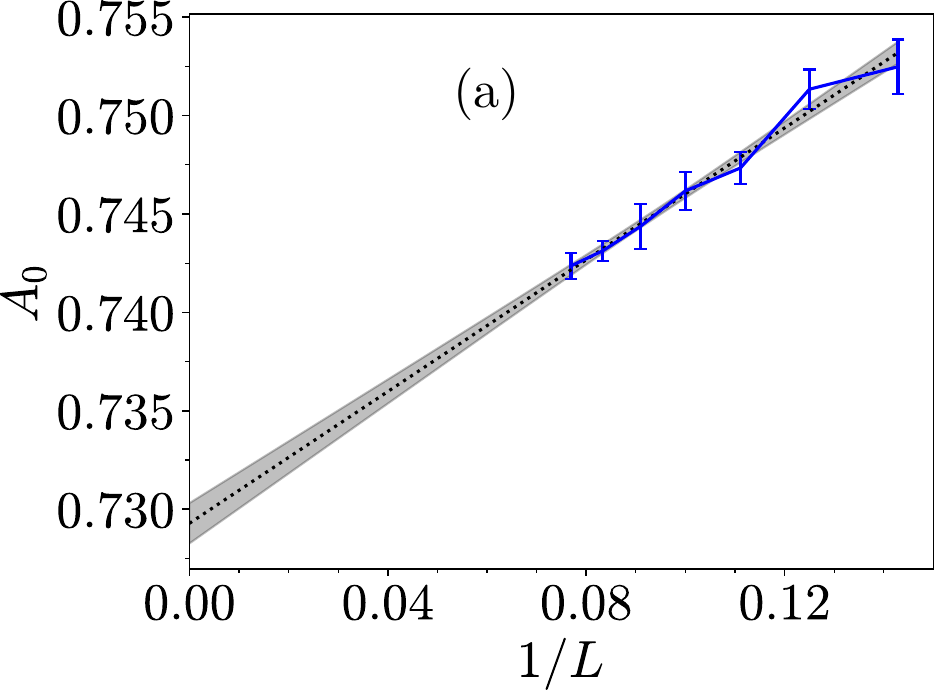}
\includegraphics [width=0.225 \textwidth,angle=0]{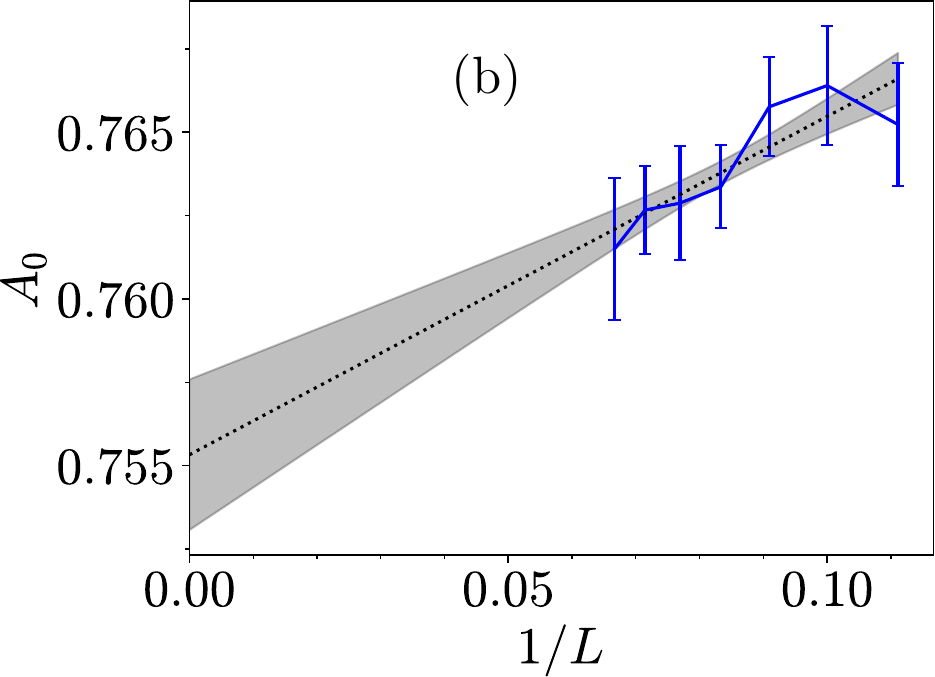}
\includegraphics [width=0.225 \textwidth,angle=0]{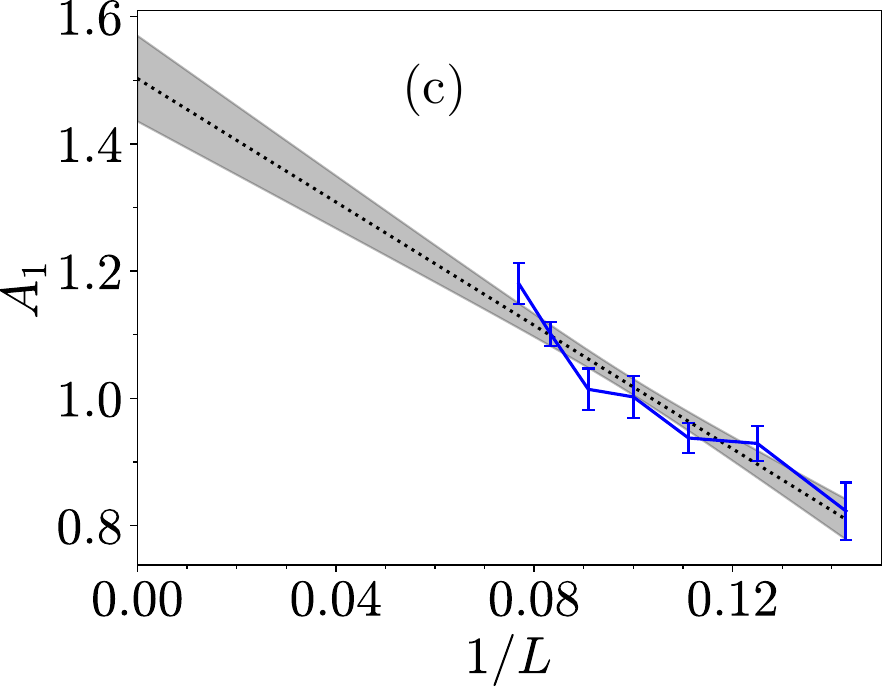}
\includegraphics [width=0.225 \textwidth,angle=0]{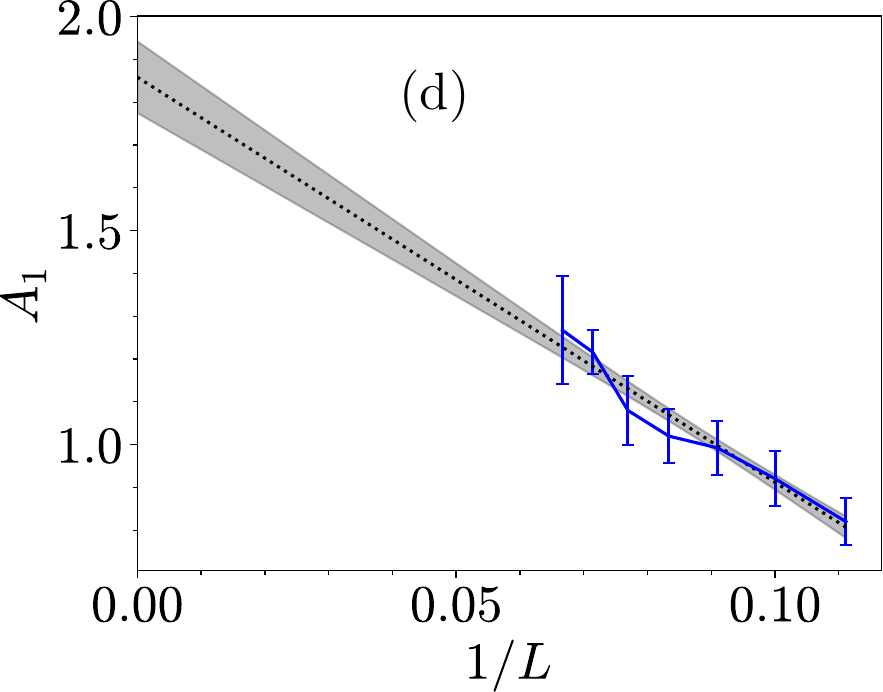}
\caption{Extrapolation of parameters $A_0$ and $A_1$ fitted to the mobility edge shape according to Eq.~\eqref{eq:ME_analytical_fit} shown with a dotted line. The gray band represents $1\sigma$ the confidence interval. The original QSM (4) shown on (a,c), while the U(1)-preserving QSM (5) is on (b,d).} \label{fig:extrapolation}
\end{figure}

From the extracted values of $\alpha_c^\varepsilon(L)$ it is clear that the critical point systematically decreases with increasing system size, although this change is small.
To obtain prediction for the critical point at the thermodynamic limit, we considered the following two parameter formula:

\begin{equation}
    \alpha_c(\varepsilon)=A_0e^{\frac{|\varepsilon-\varepsilon_0|^q}{A_1}}\;, \label{eq:ME_analytical_fit}
\end{equation}
and we fitted it to the numerically obtained values of $\alpha_c^\varepsilon(L)$ for both models. The exponent $q$ was chosen based on the form of the density of states, which was perfectly Gaussian $q=2$ for the original model, while for the U(1) symmetric version, $q=2.25$ gave the best description of the density of states. Perfect agreement of the data with those functional forms is shown in Fig.~\ref{fig:analytic_fit}.

Parameters of the fit $A_0,A_1$ are then extrapolated to the thermodynamic limit, as shown in Fig.~\ref{fig:extrapolation}. The scaling used was of the form $A_{1,2}(L)=a_{1,2}+b_{1,2}/L$, as the parameters changed almost perfectly linearly in $1/L$, where $a_{1,2},b_{1,2}$ are constants being fitted. This scaling is also well motivated analytically, as we found that both parameters $A_0,A_1$ have the leading contribution of order $\mathcal{O}(L^0)$, thus the biggest subleading contribution should be $\mathcal{O}(L^{-1})$. We also considered a scaling linear in the parameter $1/(N+L)$, which gave the same results as $1/L$, thus we are not discussing it in detail. As the shape of the mobility edge in the thermodynamic limit we took the formula~\eqref{eq:ME_analytical_fit} with parameters $A_{0}(\infty),A_1(\infty)$, while its confidence intervals were estimated using standard error propagation.

\section{Entanglement entropy of eigenstates}

In the main text, it was shown, that using entanglement entropy of eigenstates as an indicator of ergodicity breaking gives results in agreement with gap ratio analysis. At $\alpha>\alpha_c$ entropy $S$ followed volume law and the rescaled entropy $S_r$ (defined in the main text) was increased towards value $1$, as expected from the ergodic system. On the other hand, $\alpha<\alpha_c$ showed area law and $S_r$ was decreasing, clearly indicating localization. 
\begin{figure}[ht!]
\centering
\includegraphics [width=0.225 \textwidth,angle=0]{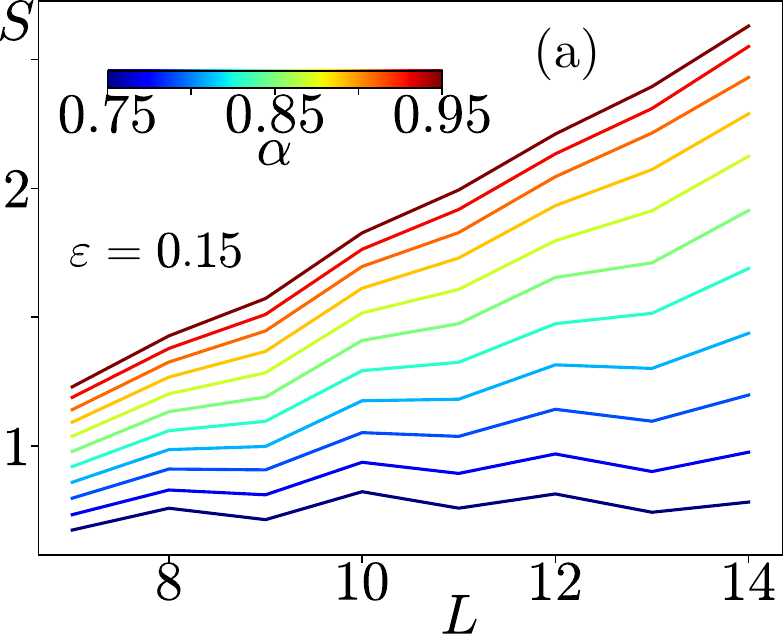}
\includegraphics [width=0.225 \textwidth,angle=0]{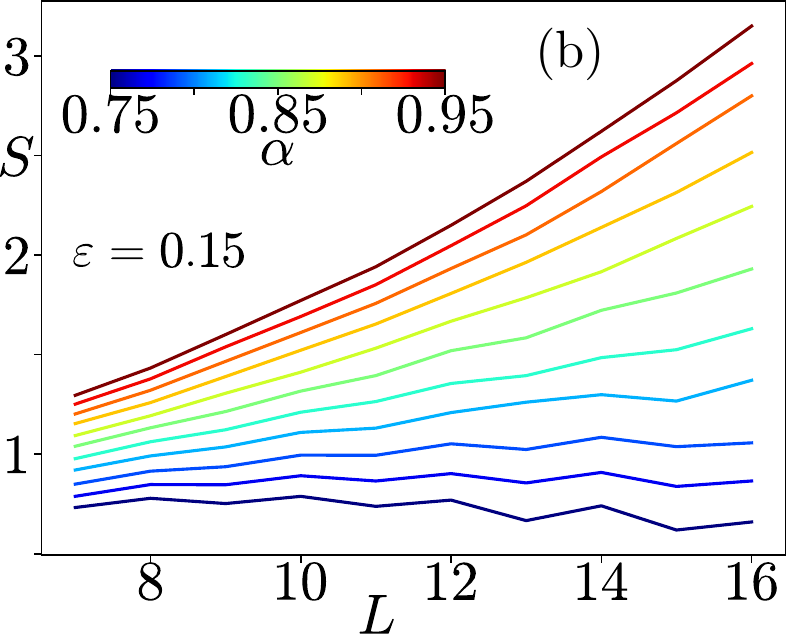}
\includegraphics [width=0.225 \textwidth,angle=0]{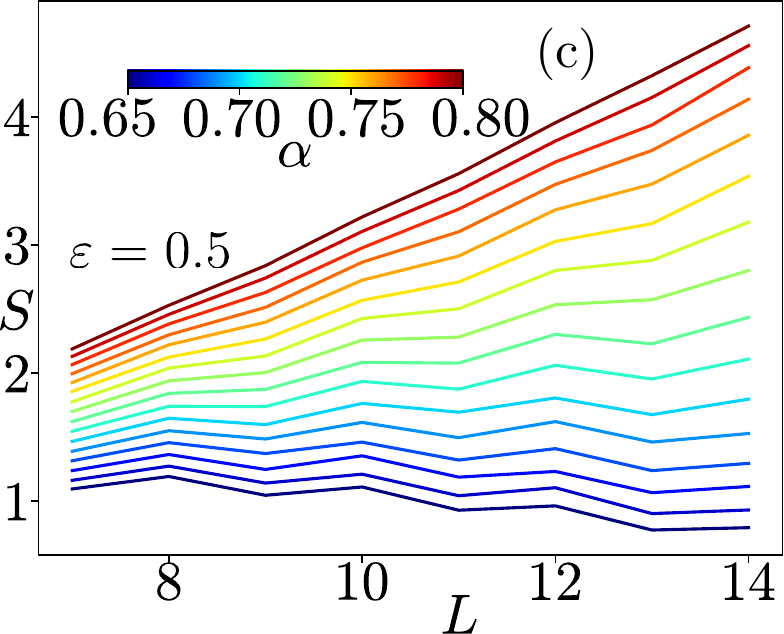}
\includegraphics [width=0.225 \textwidth,angle=0]{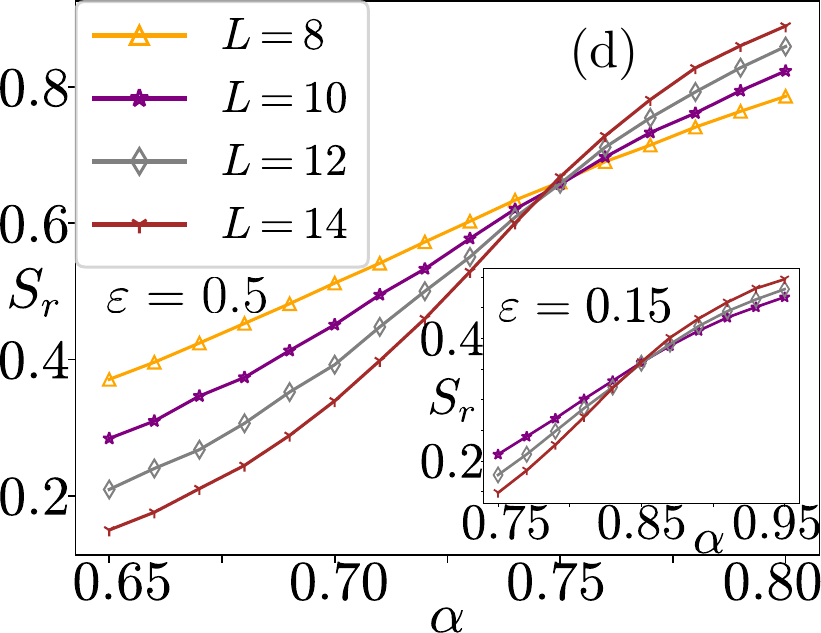}
\caption{Entanglement entropy of eigenstates in QSM. (a,b) Average entropy, $S$, far from the center of the spectrum ($\varepsilon=0.15$) in the original QSM (4) on (a), and in the U(1) variant of QSM (5) on (b). (c) Average entropy, $S$, close to the center of the spectrum ($\varepsilon=0.5$) in the original QSM (4).
(d) Dependence of the rescaled entanglement entropies $S_r$ (defined in the main text) for the original QSM (4) at $\varepsilon=0.5$. Inset shows the data for $\varepsilon=0.15$.} \label{fig:ent_supp}
\end{figure}
The decrease of entropy with increasing system size, being the consequence of $0-$dimensional nature of the system, was also understood, as the effect of going further from the GOE sun, which is the source of entropy growth in the system.

Here, we supplement those results of the main text with the equivalent of Fig.~4 for the original QSM, see Fig.~\ref{fig:ent_supp}(c,d). Moreover, we present area-volume law transition for eigenstates far from the middle of the spectrum in Fig.~\ref{fig:ent_supp}(a,b) for both studied models. The numerical procedure to obtain those plots is the same as described for entanglement entropies in the main text. The results of Fig.~\ref{fig:ent_supp} show the same behavior as the spectral analysis, with clear shift of the critical point $\alpha_c$ between area and volume laws for different values of $\varepsilon$, giving further confirmation of the mobility edge. This proves that results obtained are robust, independently of the signature of the ergodicity breaking being studied.

\section{Critical exponents and data collapse}
When the phase transition of the system is approached, the characteristic length $\xi$ diverges \cite{Cardy96} and indicators of the transition $X(L,\alpha)$ follow the scaling:

\begin{equation}
    X (L,\alpha)=f[\mathrm{sgn}(\alpha-\tilde\alpha)L^{1/\nu}/\xi],\label{eq:collapse}
\end{equation}
with a critical exponent $\nu$ and transition point $\tilde\alpha$ and $f$-continuous function. For the QSM, the characteristic length is \cite{Suntajs22}:
\begin{equation}
\xi=\frac{1}{2\left |\ln \frac{\alpha}{\tilde\alpha} \right |}.\label{eq:xi}
\end{equation}

\begin{figure}[ht!]
\centering
\includegraphics [width=0.45 \textwidth,angle=0]{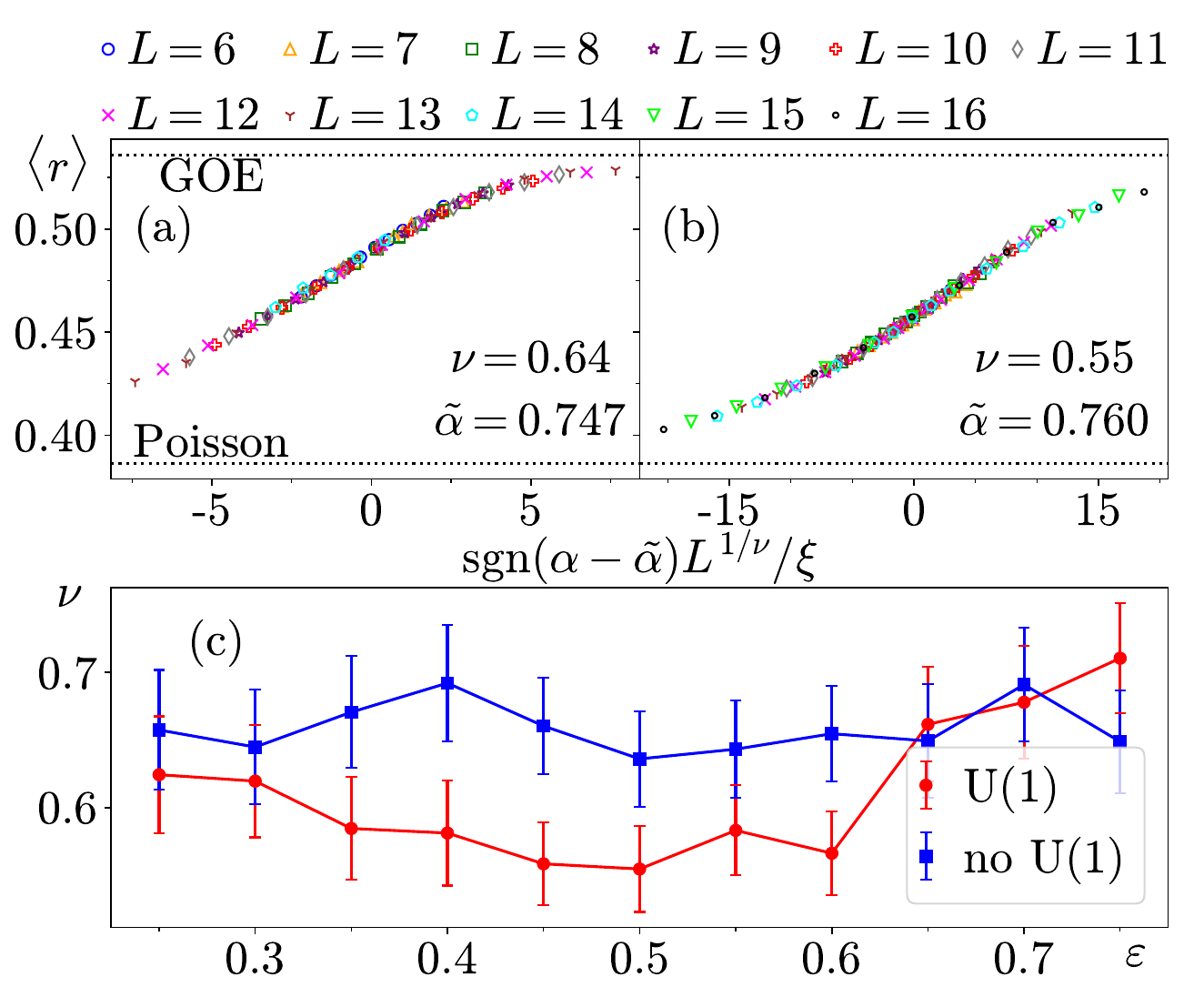}
\caption{Data collapse for (a) the original QSM for system sizes $6-14$ and (b) U(1) QSM for system sizes $7-16$. Critical parameters obtained from a cost function method are shown. Black dotted lines indicate limiting values for ergodic (GOE) and localized (Poisson) systems. (c) The critical exponent of the transition at different rescaled energies $\varepsilon$ for both models.} \label{fig:scaling}
\end{figure}

We applied the cost function method \cite{Suntajs20} to find parameters $\nu,\tilde\alpha$ that provide optimal collapse \eqref{eq:collapse} of $X=\langle r \rangle$. To this mean, we sorted $\langle r \rangle$ with respect to nondecreasing values of $\mathrm{sgn}(\alpha-\tilde\alpha)L^{1/\nu}/\xi$, obtaining an ordered list $\{X_i\}_{i=1}^{n_{\max}}$. The cost function is then given as:

\begin{equation}
    \mathcal{C}_X=\frac{\sum_{i=1}^{n_{\max}-1} |X_{i+1}-X_i|}{\max{X_i}-\min{X_i}}-1.
\end{equation}
Optimal collapse was obtained when $\alpha$ values were restricted to $\alpha\in[\alpha_*-0.05,\alpha_*+0.05]$, where $\alpha_*$ is the estimate for the transition point obtained from prior analysis, results are presented in Fig.~\ref{fig:scaling}. Critical point $\tilde\alpha$ is in a very good agreement with previously obtained $\alpha_c$.

Analogous analysis was also performed for different rescaled energies $\varepsilon$, which is summarized in Fig.~\ref{fig:scaling}c. We have found that critical exponent $\nu$ does not vary across the spectrum for both models. This confirms an intuitive expectation that the universality class of the ergodicity breaking transition in the two models does not change with the energy. It is important to note, that the value of $\nu$ depends significantly on specific choice of $\alpha$ and system sizes $L$ included in the cost function, the error associated to this effect is estimated as 5\% of $\nu$. Moreover, to calculate the statistical error of $\nu$ we repeated its calculation for $1000$ bootstrap resamples, which were created by adding to each data point $\langle r \rangle$ a random number drawn from a Gaussian distribution with $0$ mean and standard deviation equal to the uncertainty of this point. Uncertainties shown on Fig.~\ref{fig:scaling}c are a statistical combination of both types of errors. This data is consistent with the claim, that {the ergodicity breaking transition belongs to the same} universality class {in the two models}. This is further demonstrated by the fact that characteristic length \eqref{eq:xi} derived for the original QSM leads to perfect collapse of the data also for the U(1) model.

\end{document}